\documentclass[10pt, superscriptaddress, a4paper, twocolumn, showkeys]{revtex4}
\usepackage[utf8]{inputenc}
\usepackage[brazil]{babel}
\usepackage{epstopdf}
\usepackage{indentfirst}
\usepackage{graphicx,natbib}
\usepackage{amsmath,amssymb,amsfonts}

\makeatletter
\@ifundefined{textcolor}{}
{%
 \definecolor{BLACK}{gray}{0}
 \definecolor{WHITE}{gray}{1}
 \definecolor{RED}{rgb}{1,0,0}
 \definecolor{GREEN}{rgb}{0,1,0}
 \definecolor{BLUE}{rgb}{0,0,1}
 \definecolor{CYAN}{cmyk}{1,0,0,0}
 \definecolor{MAGENTA}{cmyk}{0,1,0,0}
 \definecolor{YELLOW}{cmyk}{0,0,1,0}
 }



\usepackage{babel}\usepackage{hyperref}

\makeatother

\usepackage{babel}

\makeatother

\usepackage{babel}

\begin{document}

\title{Effects of interplanetary shock inclinations on nightside auroral power intensity}

\author{D. M. Oliveira}
\email{denny@umbc.edu}
\address{NASA Goddard Space Flight Center, Greenbelt MD, USA}
\address{Goddard Planetary Heliophysics Institute, University of Maryland Baltimore County, Baltimore MD, USA}
\author{J. Raeder}
\address{EOS Space Science Center and Department of Physics, University of New Hampshire, Durham NH, USA}
\author{B. T. Tsurutani}
\address{Jet Propulsion Laboratory, California Institute of Technology, Pasadena CA, USA}
\author{J. W. Gjerloev}
\address{Johns Hopkins University Applied Physics Laboratory, Laurel MD, USA}
\address{Birkeland Centre of Excellence, University of Bergen, Norway}

\begin{abstract}
We derive fast forward interplanetary (IP) shock speeds and impact angles to
study the geoeffectiveness of 461 IP shocks that occurred from January 1995 to
December 2013 using ACE and Wind spacecraft data. The geomagnetic activity is inferred from the SuperMAG project data. SuperMAG is a large chain
which employs more than 300 ground stations to compute enhanced versions of
the traditional geomagnetic indices. The SuperMAG auroral electroject SME
index, an enhanced version of the traditional AE index, is used as an auroral
power (AP) indicator. AP intensity jumps triggered by shock impacts are correlated with both shock speed and impact angle. It is found that high AP intensity
events typically occur when high speed IP shocks impact the Earth’s magnetosphere with the shock normal almost parallel to the Sun-Earth line. This
result suggests that symmetric and strong magnetospheric compression leads to
favorable conditions for intense auroral power release, as shown previously by
simulations and observations. Some potential mechanisms will be discussed.

\end{abstract}

\maketitle



\section{Introduction}

Interplanetary (IP) shocks result from the interaction of solar disturbances
with the ambient solar wind \cite{Richter1985}. As they propagate throughout the heliosphere, IP shocks eventually interact with different bodies in the
solar system, such as planets, moons, and even asteroid5 s. In the eventual cases
in which IP shocks strike Earth, they interact with the Earth’s magnetosphere,
causing disturbances that can be detected in the near-Earth space environment, the whole magnetosphere and even the ionosphere. The first dramatic
magnetospheric effect associated with IP shock impacts is the SSC/SI$^+$ (storm
sudden commencement/sudden impulse), resulting from the sudden magnetospheric/magnetotail compression and the Earth-ward motion of the Chapman-
Ferraro current. Other effects following shock-related SSC/SI$^+$ events may also
occur: geomagnetic storms \cite{Gonzalez1999}, radiation belt perturbations \citep{Zong2009,Halford2015}, and GICs (ground-induced currents). GICs may pose risks to electric power transmission systems leading to power
grid disruptions and serious economic losses \citep{Bolduc2002,Erinmez2002,metatech-report-2010,Schrijver2014}. GICs are also associated with corrosion of pipelines and their control systems \cite{Gummow2002}.IP shocks are well known to sometimes trigger substorms as well \cite{Schieldge1970}. \par

IP shocks are well known to sometimes trigger substorms as well \cite{Schieldge1970,Kawasaki1971,Kokubun1977,Akasofu1980,Craven1986,Lui1990,Lyons1995,Lyons1996,
Zhou1999,Zhou2001,Liou2003,Yue2010,Echer2011}. In the early days before shock detection in
interplanetary space, SSCs/SI$^+$s were used to imply the impingement of an IP shock or tangential discontinuity (TD) onto the magnetosphere as an attempt
to explain geomagnetic activity following SSC/SI$^+$ events. As shown later by \cite{Kokubun1977}, who statistically examined SSC/SI$^+$ events, intense au-
roral activity always occurred when SSC/SI$^+$ amplitudes were greater than 40
nT. \cite{Smith1986} later showed that most of SSCs/SI$^+$s were caused by IP shocks rather than TDs. Precursor IMF $B_z$ events $\sim$1.5 hr prior to shock arrival have bee used to identify when shocks would be geoeffective and when
they would not be \cite{Craven1986,Zhou1999,Zhou2001,Tsurutani2003a,Yue2010,Echer2011,Liu2013} \par

Another important factor of IP shock geoeffectiveness is the IP shock impact
angle, which is the angle between the shock normal vector and the Sun-Earth
line. IP shocks with shock normals almost aligned with the Sun-Earth line
are typically driven by CMEs (coronal mass ejections) \citep{Richter1985} whereas inclined shocks tend to be driven by CIRs (corotating interaction regions) more frequently \citep{Siscoe1976,Pizzo1991}. Several studies addressing
geomagnetic activity following inclined IP shocks have been done in the past. For example, a longer than usual SSC rise time caused by
the impact of an IP shock with the Earth’s magnetosphere was observed by Wind spacecraft \cite{Takeuchi2002b}. These authors argued that
this effect should be related to the high inclination of the shock normal in the
equatorial plane. Due to its high impact angle, the IP shock took a longer
time to sweep over the magnetosphere, compressing it gradually leading to a
slow magnetospheric response. The authors suggested that simulation studies
with the impact of inclined IP shocks on the Earth’s magnetosphere should be
carried out. This suggestion was taken by other authors \cite{Guo2005} who simulated the impact of two similar IP shocks, one inclined and the other frontal, on the
Earth’s magnetosphere. They found that both systems evolved to very similar
final quasi-steady states, although in the inclined case the system took more
time to evolve in relation to the head-on case. Very similar results were found with MHD numerical simulations by other studies as well \cite{Wang2005,Samsonov2011,Samsonov2015}. A statistical study with more than 300 fast forward IP shocks was performed \cite{Wang2006} to study the same effects, i.e., the effect of IP shock inclinations on the
SSC rise time followed by IP shock impacts. They found that IP shocks impacting the Earth’s magnetosphere with shock normals almost aligned with the
Sun-Earth line caused short SSC rise times when compared to inclined IP shock  events. Their highest correlation occurred when the shocks were strong and
almost head-on. These simulation and statistical results confirmed the observation and suggestion made previously by observations \cite{Takeuchi2002b} and numerical
simulations as well \cite{Guo2005,Wang2005,Samsonov2011,Samsonov2015}. \par

Impact of IP shocks on the Earth’s magnetosphere were investigated in other simulation studies \cite{Samsonov2006a,Samsonov2007} with an MHD code especially developed to study effects
generated on the magnetosheath \cite{Samsonov2006a}. In this case, the shock
impact was frontal and the magnetospheric response was symmetric. Later, using the same MHD code,  one of the previous authors a simulated case of the interaction of a similar
IP shock, but with an inclined shock normal, with the Earth’s magnetosphere.
 He found that the inclined shock took more time to travel through the magnetosheath in comparison to his previous studies. This author argued that
inclined shocks with large downstream $v_y$ component may generate asymmetries on both dawn and dusk sides leading to non-symmetric magnetospheric
compressions. Such compressions can lead to different SSC amplitudes depending upon the side of impact. He then suggested that such effects should be
detectable by magnetometers on the ground. \par

More recently, the impact of IP shocks
with different shock normal inclinations on the Earth’s magnetosphere were simulated \cite{Oliveira2013,Oliveira2014b}. Using
the OpenGGCM MHD code \cite{Raeder2003}, these authors simulated three different cases,
namely, two inclined and one frontal. The second inclined shock was twice as
stronger as the other inclined shock, and the frontal shock had the same strength
as the first inclined shock. The shock normals of the two inclined shocks lay in
the meridian plane. The same authors reported that the head-on shock
was more geoeffective than the inclined shocks, even more geoeffective than
 the strong inclined shock. For example, the frontal shock triggered substorm
signatures and high nightside auroral energy dissipations not seen in the inclined
cases. They suggested that the frontal shock, whose shock normal was aligned
with the Sun-Earth line, compressed the magnetotail symmetrically on both
north and south sides. Such condition created an ideal scenario for the energy
 stored in the magnetotail to be leaked away and trigger auroral substorms at
Earth \citep{Oliveira2014b,Oliveira2014c}. These results suggested the same authors to look for these effects in
satellite and geomagnetic data. \par

The sequence of this simulation work was conducted by the same authors \cite{Oliveira2015a,Oliveira2015d},
who performed a statistical study of IP shock properties at 1 AU using a shock
list with events from January 1995 to December 2013. They found that the
yearly number of IP shocks is well correlated with solar activity, confirming previous observations \citep{Oh2007,Echer2003a}. Although
shocks occur more frequently during solar maxima due to the higher occurrence
of CMEs, CIRs tend to drive most shocks in solar minima. Due to the fact
that CMEs tend to drive frontal shocks and CIRs tend to drive more inclined shocks, more frontal events tend to occur during solar maxima. They also
reported that the majority of shocks found in the heliosphere at 1 AU are weak
shocks, with Mach numbers less than 3. In the same statistical study, they studied the effects of IP shock impact angles on substorm
strength indicated by an enhanced version of the AL index. They found that almost frontal shocks were generally more geoeffective than inclined shocks.
Their strongest correlation was found in the case when shocks with high speed,
or strong shocks, impacted the Earth almost frontally. \par

The goal of this paper is to study geomagnetic activity triggered by IP shocks
in correlation to IP shock speeds and impact angles. Here the geomagnetic activity is represented by auroral power (AP) intensity as inferred from an
enhanced version of the auroral electroject index AE. Correlations are obtained
from shock speed and impact angles using the same IP shock list published by a shock statistical study \cite{Oliveira2015a}, the most extensive fast forward shock study done
to date. In the following, in section 2, we present the data. In section 3, we report our results, which are summarize and briefly discussed in section 4.

\section{Data analyses}

Our data analysis is based on a list of 461 fast forward IP shocks from Wind
and ACE data from January 1995 up to December 2013 published by a recent statistical study \cite{Oliveira2015a}. The shock normal orientations were obtained from different shock normal determination methods, such as the well known magnetic and
velocity coplanarity methods \citep{Colburn1966}, and the formulas that mix plasma and IMF data 
\citep{Abraham-Shrauner1972,Abraham-Shrauner1976}. All shocks were required to satisfy the Rankine-Hugoniot conditions \cite{Tsurutani2011,Oliveira2015c}. \par

We use the SuperMAG geomagnetic station data to identify auroral power
associated with shock impingement. SuperMAG \cite{Gjerloev2009} is an international collaboration with a chain of more than 300 ground stations used to
compute the SME, SMU, and SML indices \citep{Newell2011a,Newell2011b},
the enhanced versions of AE, AU, and AL \citep{Davis1966}, respectively. The SuperMAG indices are very similar to the traditional IAGA indices:
their main difference is the fact that the former are computed based on data
of a larger number of ground stations in comparison to the latter. The SuperMAG data were obtained from the websites http://supermag.jhuapl.edu/ and
http://supermag.uib.no/. Technical issues related to SuperMAG data analysis
and assimilation were detailed in a paper entirely devoted to this subject \citep{Gjerloev2012}. \par

The SME index is used as a proxy for aurora power (AP) determinations.
This choice was based on a relation found by other authors \cite{Newell2011b}.
These authors calibrated the SME index with both Polar UVI
instantaneous images and DMSP instantaneous maps to obtain possible correlations between SME and AP. Due to time resolutions issues, the most relevant
correlation found by them was between SME and AP as determined by Polar
UVI. The linear relationship found in this previous work and used
here is:
\begin{equation}
AP = 0.048\times SME + 0.241\times(SME)^{1/2}\,,
\end{equation}
where AP is represented in GW, and the square root portion comes from the monoenergetic auroral contribution. In equation (1), AP was integrated over
the northern hemisphere polar cap between 1800-0600 magnetic local time and
60$^o$ and 80$^o$ magnetic latitude. More specifically, expression (1) indicates the
nightside AP intensity as calculated from the SuperMAG SME index. Later,
the SME index was confirmed to be the best choice to predict AP intensity instead of SMU and SML \citep{Newell2014}. \par

\begin{center}
\begin{figure}[h]
\vspace{-0.0cm}
\hspace*{0.0cm}\includegraphics[width=1.06\hsize]{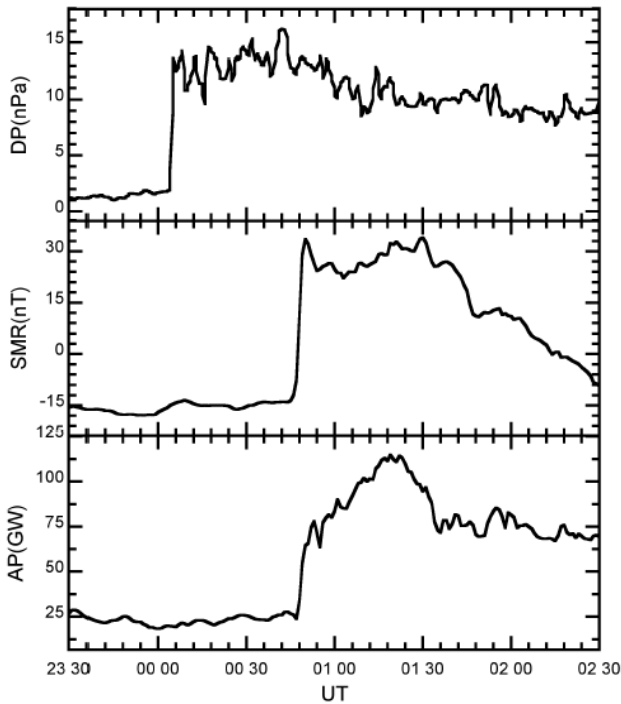}
\vspace{0.1cm}
\caption{ACE observation of an interplanetary shock on 18 April 2001 at 0005 UT and its
consequent geomagnetic activity. Top panel shows increase in dynamic pressure $\rho v^2$ (DP, in
nPa) observed by ACE. Nearly 50 minutes later, as shown in the middle panel, the Earth’s
magnetopause is struck by the IP shock, as indicated by SuperMAG ground stations with
the sharp increase in the SMR index, in nT. Finally, the bottom panel shows the increase in
AP(GW) approximately 35 minutes after shock impact.}
\end{figure}
\end{center}

The methodology used to record geomagnetic activity followed by IP shock
impacts is shown by Figure 1. This figure represents an IP shock from the
shock list compiled by a previous statistical study \cite{Oliveira2015a}. At 0005 UT on 18 April
2001, ACE observed a sharp jump in the dynamic pressure DP = $\rho v^2$ upstream
of the Earth. After approximately 50 minutes, the IP shock impinged on the
Earth’s magnetopause, and a sharp jump in SMR, the SuperMAG measurement
associated with the ring current \citep{Newell2012}, was recorded by
SuperMAG ground stations. Then, in the next 40 minutes, the SuperMAG
ground stations registered a peak in the SME index, which was used to plot AP in the last panel of Figure 1 according to equation (1). For all events in
our statical analysis, the maximum measurements in AP followed by IP shock
impacts were recorded in the time lag of 2 hours after shock impacts. If there
are more than one AP peak in this time interval, the first one is chosen as the
maximum associated with the IP shock. More details can be found in a previous work \citep{Oliveira2015a}.

\section{Results}

Figure 2 represents the statistical results obtained from our 461 IP shock
events. Figure 2(a) shows the distribution of $\theta_{x_n}$, the shock impact angle between the shock normal and the GSE Sun-Earth line. Angles close to 180$^o$ indicate that the shock normal vector is almost parallel to the Sun-Earth line. Most shocks had $\theta_{x_n}$, indicating that they typically range from moderately inclined to almost frontal shocks. Figure 2(b) represents the distribution
of shock speed vs in our shock list. The average shock speed is about 500 km/s
in the Earth’s frame of reference, and most shocks have vs below the average. This result indicates that the IP shocks observed in the heliosphere at 1 AU are
predominantly weak IP shocks with Mach numbers between 1 and 3. More details about the statistical results of the IP shocks in our database can be found
in a previous paper \citep{Oliveira2015a}.

\begin{center}
\begin{figure}
\vspace{-0.0cm}
\hspace*{0.0cm}\includegraphics[width=0.98\hsize]{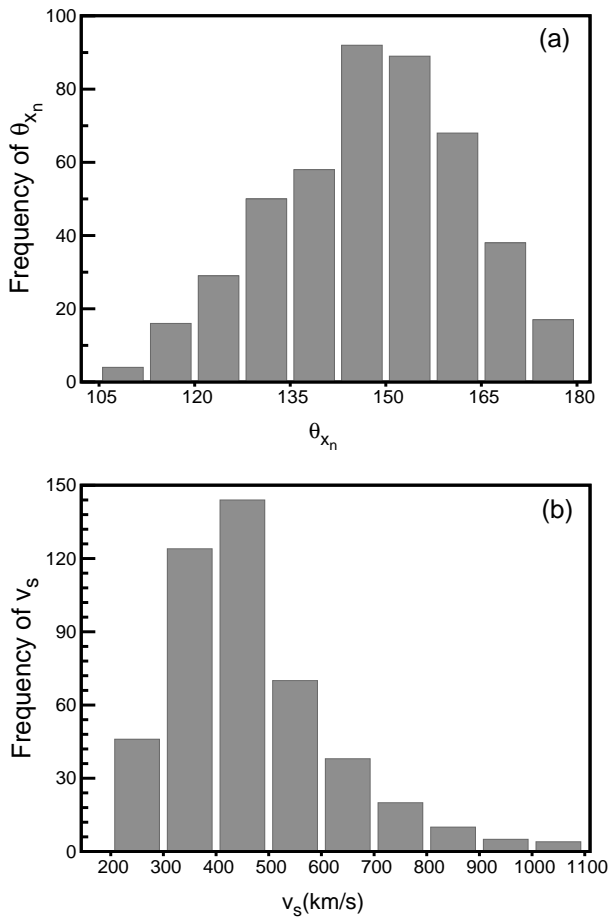}
\vspace{0.1cm}
\caption{Statistical results of fast forward IP shocks observed by Wind and ACE from January 1995 to December 2013. Figure 1(a) shows the shock impact angle distribution, and Figure 1(b) represents shock speed (km/s) distribution.}
\end{figure}
\end{center}

Correlations of variations in auroral power, $\Delta$AP, in GW, with the shock
speed $v_s$, in km/s, is shown in Figure 3. For this parameter selection, the impact
angle $\theta_{x_n}$ is held in constant intervals while the shock speed is allowed to vary.
The data are binned in three different categories: Figure 3(a), 120$^o$ $\leq$ $\theta_{x_n}$ $\leq$ 140$^o$, highly inclined shocks; 2(b), 140$^o$ $<$ $\theta_{x_n}$ $\leq$ 160$^o$ , moderately inclined
shocks; and 2(c), 160$^o$ $<$ $\theta_{x_n}$ $\leq$ 180$^o$, almost frontal shocks. Here we consider events with low auroral activity when $\Delta$AP $<$ 20 GW, and events with high
auroral activity when $\Delta$AP > 80 GW. Events with moderate auroral activity
are between these two limits. Figure 3(a) shows that most highly inclined shock
events with low auroral activity are associated with weak, low speed ($v_s$ $<$
450 km/s) shocks. Strong, high speed ($v_s$ $>$ 550 km/s) shocks are related to events with moderate auroral activity, with only one event that has low auroral
activity being caused by a strong shock. Events with moderate auroral activity
are associated with all shock strength categories with approximately the same
likelihood. There are no events with high auroral activity triggered by highly
inclined shocks in our database. The correlation coefficient in this case is R = 0.45 and the average of AP is 25.98 GW. \par

The intermediate category of shock strength has the largest number of
events, as seen in Figure 3(b). In this case, all events with low auroral activity are triggered by weak or low speed shocks. Most events with moderate
activity are associated with weak or moderate shocks. All events with high auroral activity are triggered by high speed shocks. The correlation coefficient
is R = 0.55 and $\overline{\mbox{AP}}$ = 41.37 GW. Figure 3(c) shows that all weak auroral
activity events (only three cases) are related to weak shocks. Events with moderate auroral activity are mostly associated with weak or moderate shocks, but
some are related to strong shocks. All events with intense auroral activity are
triggered by either moderate or strong shocks. The correlation coefficient R =
0.70 and the average $\overline{\mbox{AP}}$ = 64.09 GW are the highest in this category. These
results are summarized in Table 1. \par

\begin{center}
\begin{figure}
\vspace{-0.18cm}
\hspace*{0.0cm}\includegraphics[width=0.86\hsize]{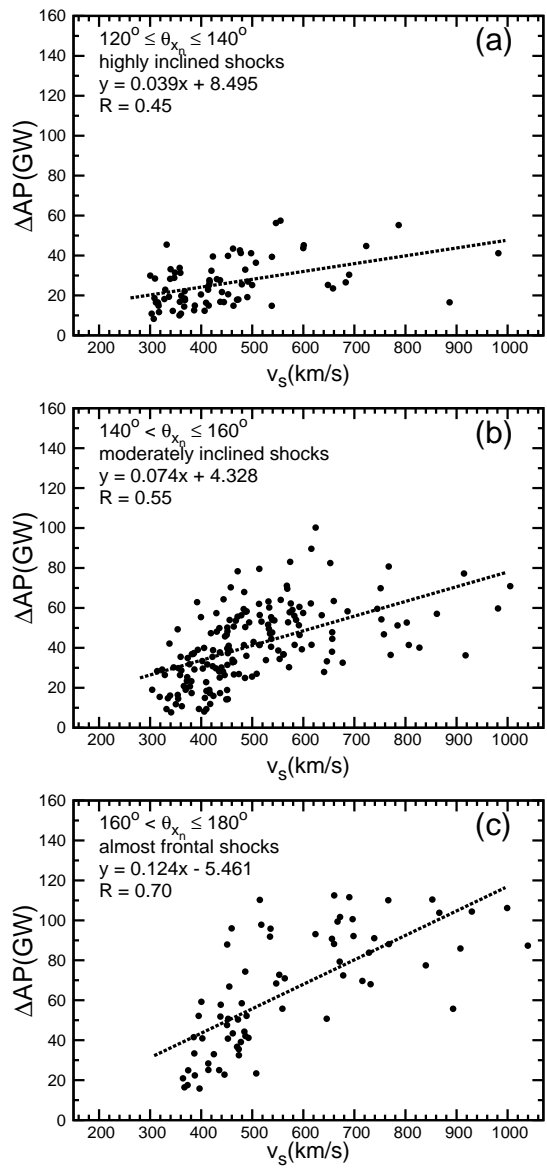}
\vspace{0.1cm}
\caption{Auroral power amplitude as a function of shock speed binned in three different shock impact angle categories: (a), 120$^o$ $\leq$ $\theta_{x_n}$ $\leq$ 140$^0$; (b) 140$^o$ $<$ $\theta_{x_n}$ $\leq$ 160$^0$; and (c), 160$^o$ $<$ $\theta_{x_n}$ $\leq$ 180$^0$
.}
\end{figure}
\end{center}

A comparison among the three cases described above shows that, on aver-
age, $\Delta$AP increases with shock speed when the impact angle is close to 180o.The correlation coefficient between the shock speed and $\Delta$AP also increases for
almost frontal shocks. \par

The opposite analysis is made in Figure 4, i.e., where the shock impact angles
are allowed to vary keeping the shock speed binned in constant intervals. The
three categories are: Figure 4(a), 300 $\leq$ $v_s$ $\leq$ 450 km/s, weak shocks; 450 $<$ $v_s$ $\leq$ 550  km/s, moderate shocks; and $v_s$ > 550 km/s, strong shocks. Figure
4(a) shows that weak shocks are associated with events with either weak or
moderate auroral activity, and are not related to events with intense auroral
activity. There are only a few weak highly inclined shocks, and most of them
cause events with moderate auroral activity. Only a few strong highly inclined shocks cause events with low auroral activity. The correlation coefficient for
highly inclined shocks, R = 0.39, and the average of 26.55 GW, are the lowest
in this case. In the category of moderate shocks, the correlation is stronger,
with R = 0.48, and the average is higher, with $\overline{\mbox{AP}}$ = 46.56 GW. There are
only a few events with low auroral activity, and most of them are triggered by highly inclined shocks and just a few by inclined shocks. Moderate almost
frontal shocks, seen in Figure 4(b), triggered either moderate or strong auroral
activity events. Typically, events with moderate auroral activity are triggered
by moderate, strong, and weak shocks. There are only a few events with high
auroral activity, and all of them are triggered by moderate almost frontal shocks. Finally, correlations for strong shock are represented by Figure 4(c). Generally,
strong shocks do not cause events with low auroral activity, with an exception
of only one event caused by a highly inclined shock. Events with moderate AP
activity are typically caused by inclined shocks, but they can also be triggered
by highly inclined or almost frontal shocks. Events with intense auroral activity
235 are caused mostly by almost frontal shocks, but a few events are caused by
inclined shocks. The correlation coefficient and AP average for strong shocks
are the highest in this category, R = 0.79 and $\overline{\mbox{AP}}$ = 62.88 GW. Table 1
summarizes the results obtained for correlations with shocks in all categories. \par

\begin{center}
\begin{figure}
\vspace{-0.18cm}
\hspace*{0.0cm}\includegraphics[width=0.86\hsize]{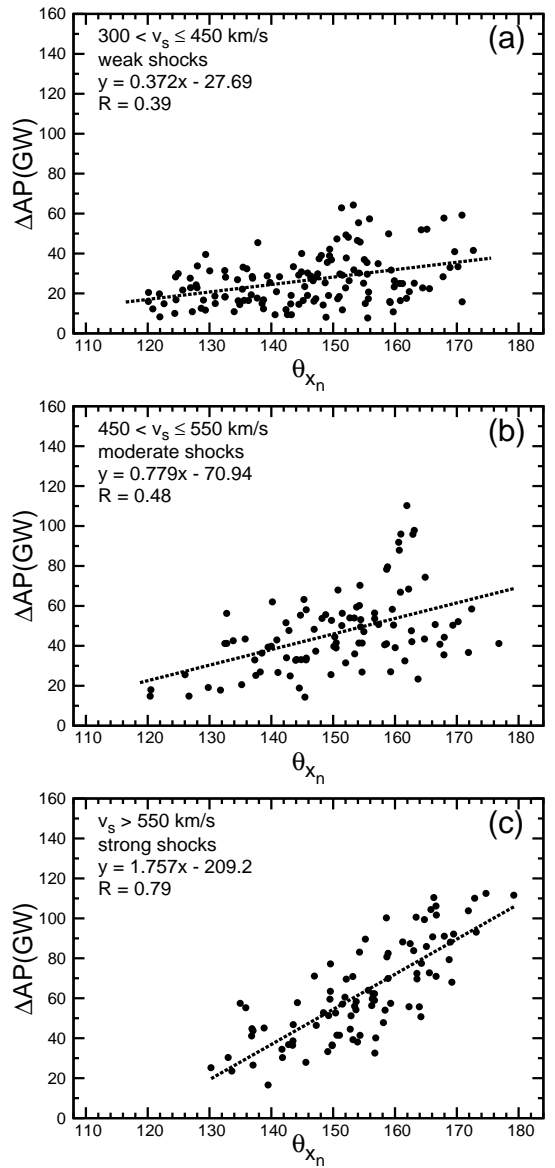}
\vspace{0.1cm}
\caption{Auroral power amplitude as a function of shock impact angle binned in three different shock speed categories: (a), 300 $\leq$ $v_s$ $\leq$ 450 km/s; (b) 450 $<$ $v_s$ $\leq$ 550 km/s; and (c), $v_s$ $>$ 550 km/s.}
\end{figure}
\end{center}

The analysis of the three panels in Figure 4 leads to a similar conclusion obtained in the analysis of Figure 3: strong, high speed shocks are generally much more geoeffective than weak slow speed shocks, and their geoeffectiveness
increases if the IP shock impacts more frontally on the Earth’s magnetosphere.
These general results were predicted previously in global
MHD simulations \cite{Oliveira2014b} and confirmed experimentally with spacecraft and geomagnetic data observations \cite{Oliveira2015a}.

\section{Summary and Conclusions}

\begin{table}
\begin{tabular}{c c c c}
\hline
\hline
\multicolumn{4}{c}{{\bf Fixed impact angle $\theta_{x_n}$, changed shock speed $v_s$}} \\
\hline
category & highly inclined & moderately inclined & almost frontal \\
R & 0.45 & 0.55 & 0.70 \\
$\overline{\mbox{AP}}$ & 25.98  & 41.37  & 64.09 \\
\hline
\hline
\multicolumn{4}{c}{{\bf Fixed shock speed $v_s$, changed impact angle $\theta_{x_n}$}} \\
\hline
 category & weak & moderate & strong \\
R & 0.39 & 0.48 & 0.79 \\
$\overline{\mbox{AP}}$ & 26.55  & 46.56  & 62.88 \\
\hline
\hline
\end{tabular}
\caption{Summary of the results obtained for the shock speed, shock impact angle, and $\Delta$AP correlation analyses.}
\end{table}

We have studied 461 fast forward interplanetary (IP) shocks using Wind
and ACE satellite data from January 1995 to December 2013. We correlated
IP shock impact angles with geomagnetic activity (auroral power intensity)
triggered by IP shock impacts. The primary result obtained here was that
high speed shocks with shock normal aligned along the Sun-Earth line (head-on
shocks) cause the greatest auroral power release. The correlation coefficient for
the cross correlation analysis in this case was 0.79, the highest of any performed
in this study. This result confirms previous numerical simulation results \citep{Oliveira2014b}, in which frontal shocks led to stronger geomagnetic activity
in comparison to the cases of inclined IP shocks. Observational results were
reported in a subsequent work \cite{Oliveira2015a}, whose authors performed a statistical analysis
correlating substorm strength and IP shock impact angles. In the case of fast
(strong) shocks, events with the strongest geomagnetic activity occurred in the cases in which the shocks impacted the magnetopause almost frontally. \par 

To explain the above results, it should be first noted that shock compression
of the magnetosphere is most effective when the inclination angle is frontal. Both
the magnetosphere and magnetotail will be compressed the most for this orientation. Greater tail lobe fields will require stronger cross tail currents to maintain them. Magnetosphere/magnetotail compression will lead to more flattened tail
closed field lines. Shock-triggering-substorm mechanisms were previously discussed by several other authors \cite{Akasofu1980,Zhou2001,Tsurutani2003a,Yue2010}. Both current disruption 
\citep{Papadopoulos1979,Lui1988,Lui1990} and magnetic reconnection 
\cite{Kokubun1977,Lui1990,Lyons1995,Lyons1996} are viable under these above conditions. \par 

The present results indicate the role of shock speed and inclination angle in
geoeffectiveness of magnetospheric energy release (auroral power). Thus this is
another factor besides magnetospheric priming that must be taken into account
in assessing auroral power release.

\vspace{0.5cm}

\begin{acknowledgments}
This work was supported by grant AGS-1143895 from the National Science Foundation and grant FA-9550-120264 from the Air Force Office of Sponsored Research. We thank the Wind and ACE teams for the solar wind data and CDAWeb interface for data availability. We thank Dr. C. W. Smith, the ACE team, and Dr. J. C. Kasper for their list compilations. For the ground magnetometer data we gratefully acknowledge: Intermagnet; USGS, Jeffrey J. Love; CARISMA, PI Ian Mann; CANMOS; The S-RAMP Database, PI K. Yumoto and Dr. K. Shiokawa; The SPIDR database; AARI, PI Oleg Troshichev; The MACCS program, PI M. Engebretson, Geomagnetism Unit of the Geological Survey of Canada; GIMA; MEASURE, UCLA IGPP and Florida Institute of Technology; SAMBA, PI Eftyhia Zesta; 210 Chain, PI K. Yumoto; SAMNET, PI Farideh Honary; The institutes who maintain the IMAGE magnetometer array, PI Eija Tanskanen; PENGUIN; AUTUMN, PI Martin Conners; DTU Space, PI Dr. J\"urgen Matzka; South Pole and McMurdo Magnetometer, PI's Louis J. Lanzarotti and Alan T. Weatherwax; ICESTAR; RAPIDMAG; PENGUIn; British Artarctic Survey; McMac, PI Dr. Peter Chi; BGS, PI Dr. Susan Macmillan; Pushkov Institute of Terrestrial Magnetism, Ionosphere and Radio Wave Propagation (IZMIRAN); GFZ, PI Dr. J\"urgen Matzka; MFGI, PI B. Heilig; IGFPAS, PI J. Reda; University of L'Aquila, PI M. Vellante; SuperMAG, PI Jesper W. Gjerloev. D.M.O. thanks the SuperMAG PI J. W. Gjerloev for the straightforward SuperMAG website for its convenience of data visualization and download.
\end{acknowledgments}


\end{document}